\documentclass[aps,prd,nofootinbib]{revtex4}
\usepackage[colorlinks=true, pdfstartview=FitV, linkcolor=blue, citecolor=red, urlcolor=magenta]{hyperref}
\usepackage{graphicx}
\usepackage{latexsym}
\usepackage{amsmath}
\usepackage{amsfonts}
\usepackage{amssymb}
\usepackage{caption}
\usepackage{graphics}
\usepackage{color}
\usepackage{epsfig}
\usepackage{graphicx}
\usepackage{subfigure} 
\usepackage{wrapfig} 
\usepackage{epstopdf}
\usepackage{verbatim}
\graphicspath{{figuras/}}

\newcommand{\be}{\begin{equation}}
\newcommand{\ee}{\end{equation}}

\begin{document}

\title{Lifshitz-scaling in CPT-even Lorentz-violating electrodynamics and GRB time delay}

\author{$^{1}$K. E. L. de Farias}
\email{klecio.limaf@gmail.com}

\author{$^{2}$T. A. M. Sampaio}
\email{thiago.muniz@ifsertao-pe.edu.br}

\author{$^{3}$M. A. Anacleto}
\email{anacleto@df.ufcg.edu.br}

\author{$^{1,3}$F.A. Brito}
\email{fabrito@df.ufcg.edu.br}

\author{$^{3}$E. Passos}
\email{passos@df.ufcg.edu.br}

\affiliation{$^{1}$Departamento de F\'{\i}sica, Universidade Federal da Para\'{\i}ba,\\ Caixa Postal 5008, Jo\~ao Pessoa, Para\'{\i}ba, Brazil.}
\affiliation{$^{2}$Instituto Federal de Educa\c{c}\~ao, Ci\^encia e Tecnologia do Sert\~ao Pernambucano - campus Salgueiro,\\ Caixa Postal 56000-000, Salgueiro, Pernambuco, Brazil.}
\affiliation{$^{3}$Departamento de F\'{\i}sica, Universidade Federal de Campina Grande,\\ Caixa Postal 10071, 58429-900, Campina Grande, Para\'{\i}ba, Brazil.}



\begin{abstract}
In this work, we consider the Ho\v{r}ava-Lifshitz scaling to rewrite higher-order derivatives in Lorentz-violating CPT-even electrodynamics. The possibility of the Lorentz invariance violation being originally suppressed by the quantum gravity energy scale $M_{\rm QG}$, may be indeed circumvented in a scenario with anisotropy between space and time above certain intermediate energy scale $\Lambda_{\rm HL} \ll M_{\rm QG}$. To test the model, we have used observational results of a cosmological time delay between high and low energy photons from gamma-ray bursts (GRB090510) event. For specific values of the Lifshitz critical exponent, we have found interesting lower bounds on the values of $M_{\rm QG}$.

\end{abstract}
 \maketitle

\section{Introduction}

The Lorentz invariance is a fundamental ingredient of the standard model, which accurately describes the properties of fundamental particles and interactions up to  {\rm TeV} scale - energy scale achieved by the Large Hadron Collider (LHC) experiment. However, the idea that this invariance can be violated at high energies, presumably in some quantum gravity energy scale $M_{QG}$ that has to be equal or around to the Planck scale ($\sim 10^{19} {\rm GeV}$) has been previously considered in several areas of theoretical physics (see, for example, Ref. \cite{liberati2013tests} and related references therein). In the present work, the Lorentz invariance violation (LIV) is introduced through higher-order derivative operators. These LIV models work as effective field theories in the ultraviolet (UV) scale regime and thus lead to changes in the dispersion relation for high energy particles \cite{Myers:2003fd} (see also Ref. \cite{Bolokhov:2007yc}).

The study of non-renormalizable terms in the LIV scenario from an effective field theory was initially proposed by Myers and Pospelov who used higher-order derivative operators coupled with a non-dynamics four-vector, $n_{\mu}$, that interact with scalar, fermionic and gauge fields \cite{Myers:2003fd}. For the electromagnetic field sector, we have, for example, the introduction of a CPT-odd
term with dimension-5 operator:
\begin{eqnarray}
S^{(5)}_{\rm MP}= - \frac{\xi}{M_{\rm QG}}\int d^4x\, n^{\alpha}F_{\alpha\delta}n^{\gamma}\partial_{\gamma}n_{\beta}\tilde{F}^{\beta\delta}.
\label{I.1}
\end{eqnarray}
Based on this idea, it is also possible to construct a CPT-even term with dimension-6 operator following the same Myers-Pospelov criteria:
\begin{equation}
S^{(6)}_{\rm MP}= - \frac{\zeta}{2M_{\rm QG}^{2}}\int d^4x\, \tilde{F}^{\mu\nu} n_{\mu}n^{\gamma}\big(n\cdot\partial\big)^{2}\tilde{F}_{\gamma\nu},
\label{I.1.1}
\end{equation}
where $\xi$ and  $\zeta$ are dimensionless parameters that control the LIV intensity over $M_{QG}$ scale suppression, $F_{\alpha\beta}=\partial_{\alpha}A_{\beta}-\partial_{\beta}A_{\alpha}$, $\tilde{F}^{\beta\delta}=\frac{1}{2}\epsilon^{\beta\delta\rho\omega}F_{\rho\omega}$, $\tilde{F}^{0i}=\frac{1}{2}\epsilon^{ijk}F_{jk}$ and $\tilde{F}_{0i}=\frac{1}{2}\epsilon_{ilm}F^{lm}$

Unlike electromagnetism modified with the original Myers-Pospelov term in Eq. (\ref{I.1}), which may violate causality due to birefringence effects that appear in its dispersion relation, it was shown in the Ref. \cite{thiago} that the use of CPT-even LIV terms as in Eq.(\ref{I.1.1}) modifying the electromagnetic sector, can result in a consistent theory (preserving causality and unitarity). Because the CPT-even terms do not produce birefringence effects in the dispersion relation, it becomes possible to perform phenomenological tests with the cosmological time delay between photons, since there is no impediment due to the different polarization modes interfering with the time of flight of the particles, as highlighted by Ref. \cite{Liberati:2009pf}.

For the future purpose of studying effects on the time of flight of the particles, we can choose a simple scenario by which we have only LIV effects on boosts transformations. The four-vector $n_{\mu} $ (which refers to the background field of the model) can be restricted to a time-like case ${(n_{\mu}\equiv1,\vec{0})}$ (this is known as isotropic models, better discussed in Refs. \cite{Kosteleck,Reyes:2010pv}).

However, both the original Myers-Pospelov modification, Eq. (\ref{I.1}), and the dimension-6  modification in Eq. (\ref{I.1.1}) can bring problems to electromagnetic field theory that are associated with the non-renormalization. To deal with these problems, we can use an intermediate energy scale $\Lambda_{HL}$ that can still maintain the anisotropy between space and time, such as in the Ho\v{r}ava-Lifshitz theory \cite{Horava:2008jf,Horava:2009uw,visser2009lorentz,passos2018}, that manages to renormalize theories that were previously non-renormalizable. In this theory, there is an explicit asymmetry between the temporal and spatial coordinates by which the Lorentz invariance is broken. This asymmetry is controlled by the critical Lifshitz exponent $z$. In order to make the coupling between the standard model and a phenomenologically relevant Lorentz-violating Effective Field Theory (EFT) possible, the Ref. \cite{pospeloveshang} shows us an upper limit of $\Lambda_{\rm HL}\le 10^{10} {\rm GeV}$.

This work aims to study the effects of Ho\v{r}ava-Lifshitz scaling on Maxwell's electromagnetism modified by the presence of LIV term given by Eq. (\ref{I.1.1}) to construct an EFT, in which their respective energy scales are parameterized to an intermediate energy scale governed by the critical exponent $z$.

Previous studies adopt certain corrections for the dispersion relations to analyze, through gamma-ray bursts (GRB) measurements, changes in photon polarization due to vacuum birefringence effects \cite{Gleiser:2001rm,Coburn:2003pi,Jacobson:2003bn,gotz2011detailed,Laurent:2011he} or to analyze the presence of time delay between the arrival of two photons due to vacuum dispersive effects \cite{Abdo2009,Chang201247,RodriguezMartinez:2006ee,Nemiroff:2011fk,Wei:2016exb,passos2017} as a method of obtaining boundaries on the LIV parameter or in the quantum gravity energy scale. We will take advantage of UV-Lifshitz behaviour to find new bounds on the LIV effects by using time delay measurements from the propagation of the electromagnetic waves due to GRB events.

This paper is organized as follows: In section \ref{II}, we have applied the Ho\v{r}ava-Lifshitz scaling process for usual Maxwell electrodynamics, proving that it preserves gauge invariance. In section \ref{sec: 2.2}, we have applied the same process in the presence of the dimension-6 LIV operator in the time-like case, analyzing its gauge invariance and getting the modified dispersion relation. In section \ref{IV}, we have studied the phenomenological aspects for this model using time-delay data in the propagation of photons seen in GRB090510 event, obtaining lower bounds for the $M_{\rm QG}$ scale. Finally, in section \ref{V}, we have presented our conclusions and perspectives.

\section{Ho\v{r}ava-Lifshitz scaling in electrodynamics}\label{II}

The Ho\v{r}ava-Lifshitz theory has also been proposed as an alternative to describing an ultraviolet (UV) behavior of general relativity \cite{Horava:2009uw}. It is known that general relativity is not UV renormalizable because its coupling constant has a negative mass dimension, i.e., $[G]=[M]^{-2}$, with $[M]$ being the mass unit. The main purpose of the Ho\v{r}ava-Lifshitz scaling is to introduce a certain asymmetry between space and time parameterized by the critical Lifshitz exponent $z$. Note that this can be approached in two ways: $i)$ The time dimension is $[t]=[L]^{z}$, and the spatial coordinate dimensions are $[\vec{r}]=[L]$; $ii)$ The time dimension is $[t]=[L]$, and the spatial coordinate dimensions are $[\vec{r}]=[L]^{z}$. Here, $[L]=[M]^{-1}$ is the length unit. Notice that in this scaling we have now $[G]=[M]^{z-3}$, which for $z=3$ makes the coupling constant dimensionless. One of the necessary conditions for renormalizability of a quantum field theory is that the coupling constant should be dimensionless or dimensionful with positive mass dimension ($z\geq3$).

The effects of these asymmetries allow us an adjustment in the dimensionality of the coupling constant, and then choose the value of the critical exponent $z$. Thus, it is expected that we will deal only with renormalizable interactions. Indeed this may happen, for example, in Ref. \cite{Horava:2009uw}, in which it was shown that $z\geq3$ is sufficient to guarantee the renormalizable character in UV scale of the general relativity. In this section, we consider how this should affect electrodynamics.

\subsection{Ho\v{r}ava-Lifshitz electrodynamics modifying the magnetic field sector} \label{ssc01}

A possibility to insert the critical Lifshitz exponent in field theories is found in Ref. \cite{pospeloveshang}. We can modify the magnetic sector of the electrodynamics and rewrite Maxwell's action as
\begin{equation}\label{c01}
S_{M,HL}=-\frac{1}{2}\int dtd^3\vec{x}\,\left[F_{0i}F^{0i}+\frac{1}{2}F_{ij}(-\Delta)^{z-1}F^{ij}\right],
\end{equation}
where $\Delta=-\partial_i\partial^i=\vec{\partial}\cdot\vec{\partial}$, the metric used is ${\rm diag} (1, -1, -1, -1 )$, and $z$, as we have aforementioned, is the Lifshitz critical exponent. When $ z > 1$, this action has an anisotropic  behavior and for $z=1$, we recover the usual theory. In principle one should include all the lower spatial derivative terms, but at large $k$ (momentum of particles), which is the limit that we are mainly interested
in the highest spatial derivative term dominates and we will keep only it.  Note that the proposed model given by Eq. (\ref{c01}) must fit the first case of anisotropy between space and time: $([t]=[L]^{z},\; [\vec{r}]=[L])$. Consequently, we have the respective dimensionality rules:
\begin{eqnarray}\label{c02}
 [\partial_t]=[L]^{-z},\;\;
 [\partial_i]=[L]^{-1}.
\end{eqnarray}
\begin{equation}\label{c03}
 [A_0]=[L]^{-\frac{1}{2}(z+1)},\;\;  [A_{i}]=[L]^{\frac{1}{2}(z-3)}.
\end{equation}
Note that, by assuming $z=1$, we recover Maxwell's theory. From here, we can define our terms with the new rescaling:
\begin{eqnarray}\label{c04}
&&t\rightarrow\Lambda_{\rm HL}^{-z+1}t,\;\;
A_0\rightarrow\Lambda^{\frac{1}{2}(z-1)}_{\rm HL}A_0,\nonumber \\&&
A_i\rightarrow\Lambda^{-\frac{1}{2}(z-1)}_{\rm HL}A_i,\;\;
\partial_t\rightarrow\Lambda^{(z-1)}_{\rm HL}\partial_t.
\end{eqnarray}
In this case, the dimension of Ho\v{r}ava-Lifshitz energy scale $\Lambda_{\rm HL}$ will follow the rule: $[\Lambda_{\rm HL}]=[L]^{-1}$, due to the rescaling process. The other quantities will have the following dimensional structures: $[t]=[L], [\partial_{t}]=[A_{0}]=[A_{i}]=[L]^{-1}$. Applying Eq.(\ref{c04}) to Eq.(\ref{c01}), we obtain
\begin{equation}\label{c05}
S_{M,HL}=-\frac{1}{2}\int dtd^3\vec{x}\,\left[F_{0i}F^{0i}+\frac{1}{2 \Lambda_{\rm HL}^{2(z-1)} }F_{ij}(-\Delta)^{z-1}F^{ij}\right].
\end{equation}
As shown in Ref.\cite{Passos:2016bbc}, this theory preserves gauge invariance.

\section{Electrodynamics with the CPT-even dimension-six operator under Lifshitz's scaling}
\label{sec: 2.2}

Based on subsec. \ref{ssc01}, where a Ho\v{r}ava-Lifshitz scaling was constructed for Maxwell's equation describing classical electrodynamics, let us now apply in the same way the scaling for the dimension-6 operator, subject to the time-like case given in Eq.\eqref{I.1.1}.  The incorporation of higher dimensional operators present some challenges such as non-unitarity/ghosts due to the higher time derivatives (see the Ref.\cite{Reyes:2010pv}). On the other hand, the purpose of HL models is to keep the same number of time derivatives and yet to improve UV behavior by including higher spatial derivatives characterizing a  Lorentz invariance violation, where this asymmetry is governed by a critical exponent z. Hence the union of both theories becomes useful to phenomenological studies in the UV regimes. Therefore, the dimension-6 Myers-Pospelov term, when rescaled, becomes:

\begin{equation}
S_{\rm MP,HL}= -\frac{\zeta}{2 M_{\rm QG}^2}\int dtd^3\vec{x}\frac{1}{\Lambda^{2(z-1)}_{\rm HL}}\tilde{F}_{0i}\partial_t^2(-\Delta)^{z-1}\tilde{F}^{0i}.
\label{2.23}
\end{equation}
A difference that is quite evident here is the fact that for the usual dimension-5 Myers-Pospelov term in Eq. \eqref{I.1} it is necessary to scale the quantum gravity mass $M_{\rm QG}$ \cite{Passos:2016bbc}, while for the dimension-6 operator, this is not the case.

So we can now write our complete model through our new rescaling. Adding the modified Maxwell's term in Eq. \eqref{c05} to that of modified CPT-even Myers-Pospelov Eq. \eqref{2.23}, we arrive at

\begin{equation}
S_{\rm MP,HL}=-\frac{1}{2}\int dtd^3\vec{x}\,\left\{\left[F_{0i}F^{0i}+\frac{1}{2 \Lambda_{\rm HL}^{2(z-1)} }F_{ij}(-\Delta)^{z-1}F^{ij}\right]+\frac{\zeta}{\Lambda^{2(z-1)}_{\rm HL} M_{\rm QG}^2}\tilde{F}_{0i}\partial_t^2(-\Delta)^{z-1}\tilde{F}^{0i}\right\},
\label{2.24.1}
\end{equation}
in which both terms have the presence of LIV. Note that, for $z=1$ limit, we recover the usual CPT-even dimension-6 in modified electrodynamics for the isotropic case, as found in the non-renormalizable standard model extension present in Ref.\cite{Kostelecky:2009zp}.
\subsection{Gauge invariance}

In order to analyze the gauge invariance of the model, we will show here that, despite the Ho\v{r}ava-Lifshitz modification, the theory remains gauge invariant. For this, we can make an analysis similar to Arnowitt-Deser-Misner's formalism in the Lifshitz gravity \cite{pospeloveshang,Arnowitt:1962hi}. Hence, by considering the decomposition of the fields $A_0$ and $A_i=A_i^T+\partial_i\varphi$, the Eq.\eqref{2.23} after some integrations by part becomes
\begin{eqnarray}
\int dtd^3\vec{x}\;\tilde{F}_{0i}\partial_t^{2}\tilde{F}^{0i}=\int dtd^3\vec{x}\;A_i^T[\partial^k\partial_t^{2}(-\Delta)^{z-1}\partial^j-\eta^{jk}\partial_t^2(-\Delta)^{z}]A_j^T.
\label{2.26.1}
\end{eqnarray}
The above result is then added to Eq. \eqref{c05}, since it is submitted by the field decomposition and can be found in \cite{Passos:2016bbc}.
\begin{eqnarray}
S^{(2)}_{\rm MP,HL}&=& \frac{1}{2}\int dtd^3\vec{x}\left\{A^T_j\left[\partial_t^2+\frac{1}{\Lambda^{2(z-1)}_{\rm HL}}(-\Delta)^{z}+\frac{1}{\Lambda^{2(z-1)}_{\rm HL}}\frac{\zeta }{ M^2_{\rm QG}}(\partial_t)^{2}(-\Delta)^z\right]\eta^{jk}A^T_k\right.\nonumber\\
&\ &\left.+\left(A_0+\dot{\varphi}\right)\Delta\left(A_0+\dot{\varphi}\right)\right\}.
\label{2.27}
\end{eqnarray}
This expression makes explicit the gauge symmetry
\begin{equation}
A_0\rightarrow A_0'=A_0+\dot{\omega}\ \ \ ;\ \ \ \varphi\rightarrow\varphi'=\varphi-\omega,
\label{c06.2}
\end{equation}
which is nothing but the original gauge symmetry: $A_{\mu}\rightarrow A_{\mu}'=A_{\mu}+\partial_{\mu}\omega$ (this is related to the fact that the Eq.(\ref{2.24.1}) is written in terms of the gauge invariant electric $F_{0i}$ and magnetic, $F_{ij}$ components).

\subsection{Dispersion relation}

Subjecting Eq. \eqref{2.27} to the gauge choice $A_0=0$, we obtain, after some integrations by parts,
\begin{eqnarray}
S^{(2)}_{\rm MP,HL}=\frac{1}{2}\int dtd^3\vec{x}A^T_j\left[\partial_t^2+\frac{1}{\Lambda^{2(z-1)}_{\rm HL}}(-\Delta)^{z}+\frac{1}{\Lambda^{2(z-1)}_{\rm HL}}\frac{\zeta }{ M^2_{\rm QG}}(\partial_t)^{2}(-\Delta)^z\right]\eta^{jk}A^T_k
\label{2.271}
\end{eqnarray}
The equation of motion is
\begin{equation}
\frac{\delta S^{(2)}_{\rm MP,HL}}{\delta A^T_j} =  \left(\partial_t^2+\frac{1}{\Lambda^{2(z-1)}_{\rm HL}}(-\Delta)^{z}+\frac{1}{\Lambda^{2(z-1)}_{\rm HL}}\frac{\zeta }{ M^2_{\rm QG}}(\partial_t)^{2}(-\Delta)^z\right)\eta^{km}A_m=0.
\label{2.28}
\end{equation}
Intending to find an equation that describes the dispersion relation for our model, we consider the following {\it Ansatz}
$$A^k(x)=\epsilon^k(p)\exp(-ik_{\mu}x^{\mu})\ \ \ \ ;\ \ \ \ k_{\mu}=(\omega,\vec{k}),$$
where $k=|\vec{k}|$. Thus, we obtain
\begin{equation}
-\omega^2+\frac{k^{2z}}{\Lambda^{2(z-1)}_{\rm HL}}-\frac{\zeta}{M^2_{\rm QG}\Lambda^{2(z-1)}_{\rm HL}}\omega^2k^{2z}=0.
\label{2.29}
\end{equation}

It is worth mentioning that in the case where $z=1$, the Lifshitz modification is absent in Eq. \eqref{2.29}, as expected. However, we still have the presence of $\mathcal{O}(\omega^4)$ modification. It occurs due to the presence of the dimension-6 operator. Finally, Eq. \eqref{2.29} leads to
\begin{equation}
\omega=\pm\frac{k^{z}}{\Lambda^{z-1}_{\rm HL}}\left(1+\frac{\zeta}{M^2_{\rm QG}\Lambda^{2(z-1)}_{\rm HL}}k^{2z}\right)^{-1/2}.
\label{2.30}
\end{equation}
It is immediately to note that in the case of $z=1$ and $\zeta=0$ we recover the relativistic dispersion relation. We see that the intensity of dispersive effects depends on the critical exponent. A preliminary observation of the Eq. \eqref{2.30} is that there is no presence of birefringence sign (that always appears in CPT-odd EFT theories). Hence it makes it possible to use this model to calculate the temporal delay in cosmological events since the time of flight of the photons does not change due to different polarization modes.
Taking the first approximation order in Eq. \eqref{2.30}, we get the following dispersion relation
\begin{equation}
\omega=\pm\left(\frac{k^{z}}{\Lambda^{z-1}_{\rm HL}}-\frac{1}{2}\frac{\zeta}{M^2_{\rm QG}}\frac{{k^{3z}}}{\Lambda^{3(z-1)}_{\rm HL}}\right).
\label{2.33}
\end{equation}
We now follow the above expression to address the phenomenological issues.
The choice of $z=1$ recover the order $\zeta(M^{-2}_{\rm QG})k^3$, where is connected with dimension-six operator. To $z=4/3$ the order of correction obtained is $\zeta(M_{\rm QG}\Lambda_{\rm HL})^{-2}k^4$ associated to operators of dimension-seven. However, by taking $z=2/3$ we found $\zeta(M^{-2}_{\rm QG}\Lambda_{\rm HL})k^2$, in which the correction order is associated with dimension-five. In the case of $z=1/3$ the correction order obtained is $\zeta(M^{-1}_{\rm QG}\Lambda_{\rm HL})^{2}k$, in which is related to operators of dimension-four. Moreover,  by taking $\zeta=0$ we recover the usual dispersion relation. Before ending this section some comments are in order. The anisotropic scaling develops a fundamental role in improving the UV behavior of non-renormalizable theories such as Einstein gravity that has fundamental coupling of the theory given by the Newtonian constant with negative mass dimension, i.e., $[G_N]=M^{-2}$. The Horava-Lifshitz gravity takes advantage of this scaling to get $[G_N]=M^{z-3}$, which for $z\geq3$ eliminates the problem of negative mass dimension. This is done in an alternative theory of gravity, with dimensions now competing in the HL action in a new fashion as $[K_{ij}]^2[d^3x][dt]/[G_N]$, where $[K_{ij}]=M^{z}$. However, suppose one considers a deformed HL gravity by introducing higher operators into the Lagrangian, say $\Delta^{z-1}$, this would produce a Newtonian constant with dimension $[G_N]=M^{3z-5}$, which now $z\geq5/3$ would also render an improved UV behavior with smaller critical exponent $z$. Although, in our setup there is no negative mass dimension coupling as in gravity theories since we are working with gauge fields, same reduction in the magnitude of $z$ occurs in a precise manner.   
\section{Phenomenological aspects: time delay between emitted photons}\label{IV}

The temporal delay occurs when emitted photons from the same source are detected at different times, being detected a more energetic photon after the arrival of a less energetic photon, resulting in a small temporal delay. According to Quantum Gravity Theory, this happens because, at high energies next to the Planck scale, space stops being continuous and becomes a discrete space, as if it would have spatial granulations.

These energetic photons may be generated from transient sources, that is, sources that give a great peak of energy in a short time. When a highly energetic photon goes through space from the source to us, it probes the discreteness of space due to its high energy, suffering a temporal delay, concerning the low energetic photon,  in its detection.

Considering the equation \eqref{2.30}, the photon velocity in our model becomes
\begin{equation}
v =\frac{\partial\omega_{+}}{\partial k} = \frac{z k^{z-1}}{ \Lambda_{\rm HL}^{z-1}} -\frac{3z\zeta}{2}\frac{\Lambda^{2}_{\rm HL}}{M^2_{\rm QG}}\frac{{k^{3z-1}}}{\Lambda^{(3z-1)}_{\rm HL}}
\label{3.6}
\end{equation}
which is a sub-luminal group velocity for photons that are under the LIV effect. 
In order to study cosmological aspects, the expression \eqref{3.6} needs to consider the universe expansion. Hence, in terms of the cosmological redshift $\mathcal{Z}^{\prime}$\footnote{Please, note the difference between two notations: $z$ will continue associated to the Lifshitz critical exponent, while $\mathcal{Z}^{\prime}$ represents the cosmological redshift.},  the equation \eqref{3.6} becomes
\begin{equation}
v(\mathcal{Z}^{\prime})= \frac{z [k(1+\mathcal{Z^{\prime}})]^{z-1}}{ \Lambda_{\rm HL}^{z-1}} -\frac{3z\zeta}{2}\frac{\Lambda^{2}_{\rm HL}}{M^2_{\rm QG}}\frac{{[k(1+\mathcal{Z^{\prime}})]^{3z-1}}}{\Lambda^{(3z-1)}_{\rm HL}},
\label{3.8}
\end{equation}
where $k (1+\mathcal{Z}^{\prime})$ is the photon momentum when it was emitted from the source located at redshift $\mathcal{Z}^{\prime}$.
The light-travel distance, considering the universe expansion in a $\Lambda\mbox{CDM}$ Universe, is given by \cite{RodriguezMartinez:2006ee}:
\begin{equation}
x(t)=\frac{1}{H_0} \int^{t}_{0} v(t^{\prime})dt^{\prime}= \frac{1}{H_0} \int_0^\mathcal{Z} v(\mathcal{Z}^{\prime})\frac{d\mathcal{Z}^{\prime}}{(1+\mathcal{Z}^{\prime})\sqrt{\Omega_{\Lambda}+\Omega_m(1+\mathcal{Z}^{\prime})^3}},
\label{3.81}
\end{equation}
where $H_0\simeq 70 \mbox{km/s/Mpc}=1.505\times10^{-42}\rm GeV$ is the Hubble constant, while $\Omega_{\Lambda} \simeq 0.685$ and $\Omega_m \simeq 0.315$ are the mass density and the dark energy density parameters, respectively.

By using the Eq. \eqref{3.8} into Eq. \eqref{3.81} we find
\begin{equation}
x(\mathcal{Z})= \frac{1}{H_0} \int_0^{\mathcal{Z}} \left\{ \frac{z [k(1+\mathcal{Z^{\prime}})]^{z-1}}{ \Lambda_{\rm HL}^{z-1}} -\frac{3z\zeta}{2}\frac{\Lambda^{2}_{\rm HL}}{M^2_{\rm QG}}\frac{{[k(1+\mathcal{Z^{\prime}})]^{3z-1}}}{\Lambda^{(3z-1)}_{\rm HL}}\right\}\frac{d\mathcal{Z}^{\prime}}{(1+\mathcal{Z}^{\prime})\sqrt{\Omega_{\Lambda}+\Omega_m(1+\mathcal{Z}^{\prime})^3)}}.
\label{3.11}
\end{equation}
The Eq. \eqref{3.11} corresponds as well to the time of flight of the photons under the effects of LIV induced by the considered model. In the low/intermediary energy: $\Lambda_{\rm HL} \sim {\rm IR\,regime}$, the second term becomes negligible. Thus, the photon travel time recovers the usual one in the standard relativistic cosmology when we assume $z=1$ (returning to Maxwell's general results too). The first term of Eq. \eqref{3.11} is associated with the photon time of flight under a not so high energy scale (many orders lower than the quantum gravity scale), can be motivated by a suppression under the Lifshitz energy scale $\Lambda_{HL} \ll M_{QG}$ where would induce a weak anisotropy at space-time. Consequently, the time difference between a high-energy and a low-energy photon in a Ho\v{r}ava-Lifshitz LIV scenario is\footnote{For more details of cosmological time delay formula derivation due to LIV effects, see Ref \cite{Jacob:2006gn}.}
\begin{equation}
|\Delta t |\approx\frac{3z}{2}\zeta\frac{\Lambda^{3(1-z)}_{\rm HL}}{M^2_{\rm QG}}H_0^{-1}(k_h^{3z-1}-k_l^{3z-1})\int_0^\mathcal{Z} \frac{(1+\mathcal{Z}^{\prime})^{3z-2} d\mathcal{Z}^{\prime}}{\sqrt{\Omega_{\Lambda}+\Omega_m(1+\mathcal{Z}^{\prime})^3}},
\label{3.13}
\end{equation}
where $k_h$ and $k_l$ represents the higher energy and lower energy of the photons, respectively.
Starting from the fact that LIV effects would occur just under $M_{QG}$ scale, we can consider $\zeta = 1$ to estimate the quantum gravity energy. Using upper bounds for phenomenological values of $\Delta t_{\rm LIV}$, we can obtain lower bounds for $M_{QG}$, that is
\begin{equation}
M_{\rm QG}=\sqrt{\frac{3z}{2}\zeta\frac{\Lambda^{3(1-z)}_{\rm HL}}{\Delta t}H_0^{-1}(k_h^{3z-1}-k_l^{3z-1})\mathcal{I}(\mathcal{Z})},
\label{3.14}
\end{equation}
where $\mathcal{I}(\mathcal{Z})$ is the integral in Eq. \eqref{3.13}.

To estimate this bound, we will use data from GRB090510, which was studied in Ref. \cite{Abdo2009}. This GRB proved to be a promising event for the analysis of dispersive effects due to the highly energetic photons were detected on the GeV scale. The source was located at a redshift $\mathcal{Z} = 0.90$. Approximately $859 \mbox{ms}$ after the initial emission around $100$ keV, a peak of $31 \mbox{GeV}$ was detected (this corresponds to the most conservative LIV time delay choice). With these data, we can estimate, for different values of the critical exponent $z$,  lower bounds for $M_{QG}$, as can be seen in the table below.
\begin{table}[h]
\centering
\label{t1}
\caption{Lower bounds on $M_{QG}$ in terms of particular values of Lifshitz critical exponent $z$.}
\begin{tabular}{|l||l|l|l|}
\hline
\multicolumn{1}{|c||}{\ } & z=1/2  & z=2/3   & z=1            \\ \hline\hline
$M_{\rm QG}(\mbox{GeV})$                             & $5.93\times10^{16}$  & $4.72\times10^{14}$ & $2.71\times10^{10}$  \\ \hline
\end{tabular}
\end{table}
The Lifshitz energy scale used to arrive at that result is $\Lambda_{HL}=10^{10} \mbox{GeV}$ assuming that is a lower bound, according to Ref.\cite{pospeloveshang}. When we choose $z=1$, we eliminate the contribution arising from the Ho\v{r}ava-Lifshitz modification. Therefore, we end up with the pure LIV contribution of a dimension-6 CPT-even term, which brings quadratic dispersive effects. The lower bound for $M_{\rm QG}$ in this case, which is of the order of $10^{10}\mbox{GeV}$, is the same found in the literature for this type of dimension-6 EFT.

For any value of $z \neq 1$, the Lifshitz modification will contribute to the result, as it will count the presence of the intermediate energy scale $\Lambda_{\rm HL}$. For $z=2/3$, for example, we have an analogous case to what would be a linear order dispersive effects (from a pure Myers-Pospelov term of dimension-5, for example). It is interesting to note that the lower limit of $M_{QG}$ does not exceed Planck's mass, unlike the results found in the literature that consider a usual dispersion at a linear level.

\begin{figure}[h]
\begin{center}
\includegraphics[scale=0.7]{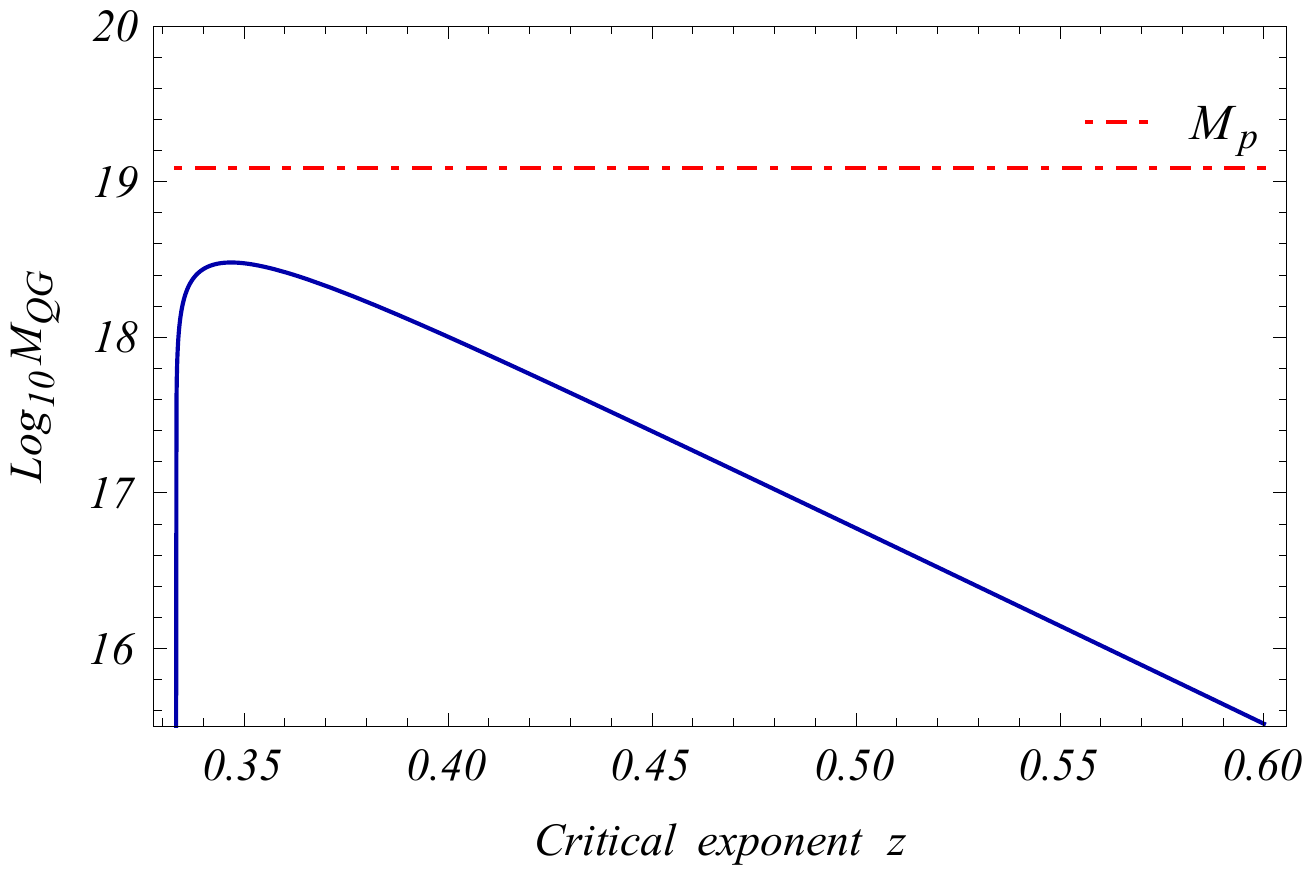}
\caption{The behavior of constraints on $M_{QG}$ in terms of Lifshitz critical exponent $z$.}
\label{g1}
\end{center}
\end{figure}

From Eq. \eqref{3.14}, we have plotted a graph to observe the behavior of constraints on $M_{QG}$ as function of $z$ as it varies from $0$ to $3$. The line on the top of graph \ref{g1} corresponds to the Planck mass, i.e., $M_{\rm P}=1.22\times10^{19}\mbox{GeV}$.

The case $z=1/3$ and does not represent any physical relevance since the Eq. \eqref{3.14} there is no energy dependence, canceling dispersive effects. For $z<1/3$, we get imaginary results. Then we should consider the constraints to $z>1/3$. Furthermore, the values next to $z=1$ produce weak lower bounds of $M_{\rm QG}$. However, it is interesting to see that the function does not exceed the Planck energy scale $M_{\rm P}$ under any circumstances.

For $z\rightarrow 1/3$, we get some interesting behavior. Normally the low energy $k_l$ is discarded in some models for being many orders smaller than the high energy value $k_h$, but in our model, it has an interesting role. This happens because $k_l^{3z-1}$ assumes values close to $k_h^{3z-1}$ in the Eq. \eqref{3.14}, resulting in a reduction in the graph growth and bringing up the $M_{\rm QG}$ to values near to $M_{\rm P}$. There is a maximum $M_{\rm QG}$ value close to $z=1/3$ before the graph starts to decay. This particular value depends on the energy difference between the photons. It is worth to highlight that the acceptable physical values lie in the interval $1/3<z\leq1$, however, as can be seen in the graphic \ref{g1} the best results are the values in the interval of $z=1/3$ and $z=1/2$, which characterizes a presence of intermediate energy between operators of dimension four and five. This result is possible due to the introduction of critical exponent $z$, where we can get control of the energy parameter. The peak in this graph is obtained when $z=0.347$, i.e., $M_{\rm QG}\ge 3.02\times10^{18} \mbox{GeV}$, which is a strong lower bound, since it is relatively close to $M_{\rm P}$.

\section{Conclusions}\label{V}

In this work, we have analyzed the LIV of a Lifshitz-scaling to a CPT-even Maxwell-Myers-Pospelov model, where the Lifshitz-type gauge-invariant of usual electrodynamics can be recovered. The dispersion relation of the theory was computed, showing a vacuum dispersive behavior with a dependence on the Lifshitz critical exponent $z$, which induces some anisotropy between space and time near to an intermediate energy scale $\Lambda_{\rm HL}$. There is no appearance of vacuum birefringence, which allowed us to test this model by analyzing the time delay between photons considering cosmological distances.

Finally, we have obtained lower bounds for the $M_{\rm QG}$ for any physical relevant Lifshitz critical exponent related to our model. These limits were determined using the observational results of the gamma-ray burst GRB090510. For $z = 1$, which corresponds to the usual dimensional-6 operator, we found the same results already obtained in the literature for quadratic order dispersion ($\ge 10^{10} \mbox{GeV}$). For $z = 2/3$, that is related to a linear dispersion, was shown that the bound for $M_{\rm QG}$ recovers a realistic scenario where LIV may be possible ($\ge 10^{14} \mbox{GeV}$, i.e., below Planck's mass). This result is an improvement from the outcome found in the literature where the Ho\v{r}ava-Lifshitz scaling is absent, which predicts bounds beyond Planck's mass. Finally, the greater difference in energies of the two considered photons was found for values of $z \rightarrow 1/3$, that is where the bounds of $M_{\rm QG}$ tend to approximate to the Planck's mass ($\ge 10^{18} \mbox{GeV}$). Therefore, this can be relevant for further phenomenological investigations and opening a window for future researches in order to understand the role of LIV in the standard model considering a quantum gravity theory, such as Ho\v{r}ava-Lifshitz gravity. We also emphasize that this behavior of Planck energy as a function of $ z $, differs from what was observed in Ref.\cite{Passos:2016bbc} which show that the upper bounds for the time delay derived from a Lifshitz-scaling to a CPT-odd Maxwell-Myers-Pospelov model is not much sensitive with respect to the critical exponent $z$.

{\acknowledgments}

We would like to thank CNPq, CAPES and PRONEX/CNPq/FAPESQ-PB (Grant no. 165/2018), for partial financial support. MAA, FAB and EP acknowledge support from CNPq (Grant nos. 306962/2018-7, 312104/2018-9, 304852/2017-1).





%


\end{document}